\documentclass{caosp306}

\usepackage{natbib}
\usepackage{aas_macros}
\usepackage{graphicx}
\graphicspath{{images/}}

\usepackage{bm}
\usepackage{amsthm}
\usepackage{amsmath}
\usepackage{amssymb}
\usepackage{amsfonts}

\newcommand{\pc}{\mathrm{pc}}

\newcommand{\msun}{M_\odot}

\newcommand{\trel}{t_\mathrm{rh}}

\newcommand{\tauv}{\tau_\mathrm{v}}
\newcommand{\nbin}{n_\mathrm{bin}}

\articleNo{000}
\pubyear{2019}
\volume{XX}
\volnumber{YY}
\firstpage{1}
\received{DATE, 2019}
\accepted{DATE, 2019}

\begin{document}

\htitle{Primordial mass segregation of star clusters: The role of binary stars}
\hauthor{V.\,Pavl\'ik}
\title{Primordial mass segregation of star clusters: The role of binary stars}
\author{V\'aclav Pavl\'ik \inst{1,} \inst{2,} \inst{3}}

\institute{Astr. Inst. of Charles University, V Hole\v{s}ovi\v{c}k\'ach 2, 180~00~Prague~8, CZE
\and Observatory and Planetarium of Prague, Kr\'alovsk\'a obora 233, 170~21~Prague~7, CZE
\and Astronomical Institute of the Czech Academy of Sciences, Bo\v{c}n\'i~II~1401, 141~31~Prague~4, CZE
\email{pavlik@asu.cas.cz}}

\date{October 31, 2019}

\maketitle

\begin{abstract}
Observational results of young star-forming regions suggest that star clusters are completely mass segregated at birth. As a star cluster evolves dynamically, these initial conditions are gradually lost. For star clusters with single stars only and a canonical IMF, it has been suggested that traces of these initial conditions vanish at $\tauv$ between 3 and 3.5 half-mass relaxation times.
By the means of numerical models, here we investigate the role of the primordial binary population on the loss of primordial mass segregation. We found that $\tauv$ does not seem to depend on the binary star distribution, yielding $3 < \tauv / \trel < 3.5$.
We also conclude that the completely mass segregated clusters, even with binaries, are more compatible with the present-day ONC than the non-segregated ones.
\keywords{methods: numerical, data analysis -- star clusters: individual (ONC) -- stars: formation, binaries}
\end{abstract}

\defcitealias{pav_segr}{Paper~I}

\section{Introduction}

Mass segregation is a prominent feature present in evolved star clusters due to their dynamical evolution \cite[e.g.][]{chandrasekhar,chandra_vonNeumann1,chandra_vonNeumann2} but not only there. Recent \textsl{ALMA} observations of the Serpens South star-forming region by \citet{plunkett} suggest that young clusters are born completely mass segregated. Despite observing a~general tendency of clusters to evolve towards higher mass segregation, it may both increase and decrease due to two-body encounters that lead to energy equipartition. In \cite[][hereafter Paper I]{pav_segr}, we were the first to point out that the degree of mass segregation of a non-segregated and a completely segregated system is gradually settled at a similar level, their primordial differences vanish and both initial conditions become observationally indistinguishable after some time designated as $\tauv$. Based on our numerical $N$-body models with single stars, we estimated this time to $3 < \tauv / \trel < 3.5$ \citep[where $\trel$ is the half-mass relaxation time; cf.][]{spitzer_hart_relax}. Most (if not all) stars are preferentially born in binary systems \citep[e.g.][]{kroupa95a,go_kr05} -- 42\,\% of field (i.e.\ old) M-dwarfs \citep{fi_ma92}, 45\,\% of K-dwarfs \citep{may_etal92} or 57\,\% of G-dwarfs \citep{duq_may91,rag_etal10} are reported in binaries, and the binary fraction increases with the stellar mass. Hence, for this conference contribution, we extend the work of \citetalias{pav_segr} by studying the evolution of mass segregation in star clusters that include primordial binaries.

\section{Models}

We evolved several realisations of $N$-body models with 2.4k stars \citep[comparable number to the Orion Nebula Cluster, ONC,][]{onc_data} and with the \citet{kroupa} IMF for several relaxation times using \texttt{nbody6} \citep{aarseth}. 
For each model, we used two extreme primordial mass segregations according to a~method of \citet{baumgardt_segr} -- none or complete.

In all models, we injected a conservative 50\% binary fraction initially (i.e.\ 601 binary stars in total), while the binary pairing was drawn from a uniform distribution of mass ratio ($0.1<q<1.0$) in the mass range above $5\,\msun$ and was random for the remaining stars up to the desired percentage \citep[cf.][]{mcluster} -- the model is labelled \texttt{P:uni}.
The semi-major axes were distributed according to \citet{sana_etal12} and \citet{oh_etal15} period distributions for stars with $m > 5\,\msun$ and according to \citet{kroupa95a} for lower-mass stars.
Eccentricity distribution of high-mass systems is taken from \citet{sana_evans11} and is thermal for low-mass stars \citep[cf.][]{heggie,duq_may91,kroupa08}.

\section{Results}

\begin{figure}[!t]
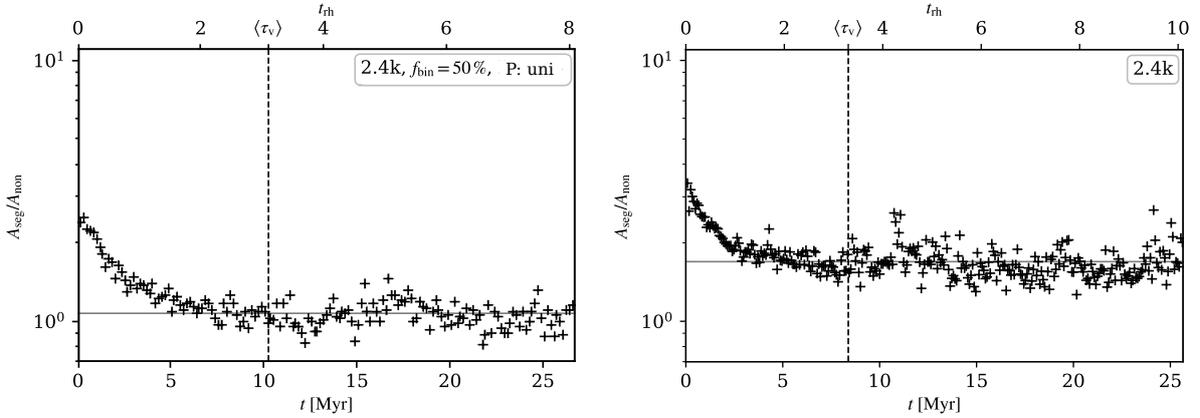

	\centering
	\hspace*{-.17\linewidth}
	\hbox{
	\includegraphics[width=.65\linewidth]{{{images/bin_seg/uni_seg_par_bin_0.1-10.0}}}
	\includegraphics[width=.65\linewidth]{{{images/paperI/m1318sun_seg_par_bin_0.1-10.0}}}
	}
	\caption{Evolution of the ratio given by Eq.~\eqref{eq:area} in time. The dashed line and the value $\langle\tauv\rangle$ represent the mean time when the slope of the data points became flat. The corresponding horizontal slope is plotted by a grey line. The left plot presents a model with initial binary population, the right plot is taken from \citetalias{pav_segr} for comparison.}
	\label{fig:tauv}
\end{figure}

\begin{figure}[!t]
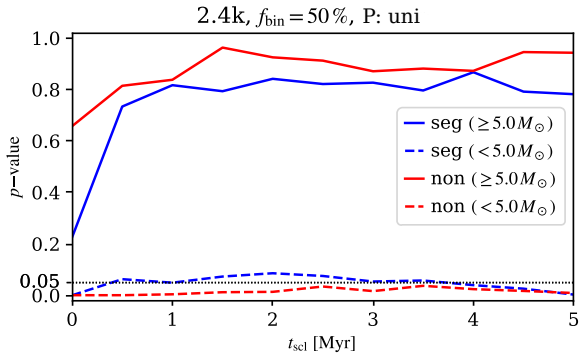

	\centering
	\includegraphics[width=.65\linewidth]{{{images/bin_seg/uni_onc_ks_extscl_m5_2.5}}}
	\caption{Results of the KS test between the ONC data and our model with 50\,\% binaries. Only the model with scaling and extinction is shown.}
	\label{fig:onc}
\end{figure}

Our clusters with binaries evolve in a similar fashion to the single star models presented in \citetalias{pav_segr} (compare the plots in Fig.~\ref{fig:tauv}). The primordially fully mass segregated clusters lose their initial ordering gradually before settling at some level of mass segregation. Clusters without initial mass segregation establish it dynamically and \emph{again} settle almost at the same level. As in \citetalias{pav_segr}, we investigate the evolution of mass segregation using the spatial distribution of mean mass and the integral parameter $A$, i.e.\ for a $k$-th bin at radius $r_k$
\vspace{-5pt}
\begin{equation}
	\label{eq:area}
	A = \sum_{k = 1}^{\nbin}{\frac{\langle m(r_k) \rangle}{\Delta r_k}} \,,
	\text{ with }
	\langle m(r_k) \rangle = \frac{\sum_{i = 1}^k{m_i}}{\sum_{i = 1}^k{n_i}} \,,
	\vspace{-5pt}
\end{equation}
where $n_i$ and $m_i$ are the number of stars and their total mass in an $i$-th bin, respectively, $\Delta r_k$ is the width of the $k$-th bin and $\nbin$ is the total number of bins (bins here are logarithmically equidistant). In particular, $r_1 = 0.1\,\pc$, $r_{\nbin} = 10\,\pc$ and $\nbin = 50$.

The time when the difference of initial conditions vanishes seems independent on the initial binary star fraction. Systems with 50\,\% binaries (\texttt{P:uni}) have $3 < \tauv / \trel < 3.5$ (see the left panel of Fig.~\ref{fig:tauv}) which is equivalent to the systems of similar population with single stars only from \citetalias{pav_segr} (see the right panel of Fig.~\ref{fig:tauv}).

We also tested whether the initial conditions with binaries are still compatible with the observed ONC. In the case of the primordially mass-segregated models, those where elongation (scaling) of the cluster and extinction was accounted for (cf. Sect.~4 in \citetalias{pav_segr}) have the KS test $p>0.05$ at the time which is equivalent to the current age of the ONC, i.e.\ 2.5\,Myr (see Fig.~\ref{fig:onc}). In the case of the initially non-segregated models, none is compatible with the present-day ONC, not even with scaling and extinction.

\section{Conclusions}

This conference contribution is a follow-up of the work of \citet[][Paper I]{pav_segr}. We have started to investigate the role of a primordial binary star population on mass segregation in star clusters of the size of the ONC.

In the models with 50\,\% initial binary stars, the mean time when the primordially mass segregated and the non-segregated models became indistinguishable was $\langle\tauv\rangle \approx 3.2\,\trel$, i.e.\ comparable to the single star models presented in \citetalias{pav_segr}.

We have also compared our models with the present-day Orion Nebula Cluster. The only compatible model is the one with primordial mass segregation if we also account for interstellar extinction and scaling of the ONC, as presented in \citetalias{pav_segr}.

\acknowledgements
This study was supported by Charles University through grant SVV-260441 and by the Czech Science Foundation through the project of Excellence No.~19-01137J.
Computational resources were provided by the CESNET LM2015042 and the CERIT Scientific Cloud LM2015085, provided under the programme ``Projects of Large Research, Development, and Innovations Infrastructures''.
The author also greatly appreciates comments from Pavel Kroupa.

\bibliographystyle{caosp306}
\bibliography{pavlik}

\begin{thebibliography}{22}
\expandafter\ifx\csname natexlab\endcsname\relax\def\natexlab#1{#1}\fi

\bibitem[{Aarseth(2003)}]{aarseth}
Aarseth, S.~J. 2003, {\it Gravitational N-Body Simulations} (Cambridge, UK:
  Cambridge University Press)

\bibitem[{{Baumgardt} {et~al.}(2008){Baumgardt}, {De Marchi}, \&
  {Kroupa}}]{baumgardt_segr}
{Baumgardt}, H., {De Marchi}, G., \& {Kroupa}, P., {Evidence for Primordial
  Mass Segregation in Globular Clusters}. 2008, {\it \apj}, {\bf 685}, 247,
  DOI: 10.1086/590488

\bibitem[{{Chandrasekhar}(1943)}]{chandrasekhar}
{Chandrasekhar}, S., {Dynamical Friction. I. General Considerations: the
  Coefficient of Dynamical Friction.} 1943, {\it \apj}, {\bf 97}, 255, DOI:
  10.1086/144517

\bibitem[{{Chandrasekhar} \& {von Neumann}(1942)}]{chandra_vonNeumann1}
{Chandrasekhar}, S. \& {von Neumann}, J., {The Statistics of the Gravitational
  Field Arising from a Random Distribution of Stars. I. The Speed of
  Fluctuations.} 1942, {\it \apj}, {\bf 95}, 489, DOI: 10.1086/144420

\bibitem[{{Chandrasekhar} \& {von Neumann}(1943)}]{chandra_vonNeumann2}
{Chandrasekhar}, S. \& {von Neumann}, J., {The Statistics of the Gravitational
  Field Arising from a Random Distribution of Stars II}. 1943, {\it \apj}, {\bf
  97}, 1, DOI: 10.1086/144487

\bibitem[{{Duquennoy} \& {Mayor}(1991)}]{duq_may91}
{Duquennoy}, A. \& {Mayor}, M., {Multiplicity among solar-type stars in the
  solar neighbourhood. II - Distribution of the orbital elements in an unbiased
  sample.} 1991, {\it \aap}, {\bf 500}, 337

\bibitem[{{Fischer} \& {Marcy}(1992)}]{fi_ma92}
{Fischer}, D.~A. \& {Marcy}, G.~W., {Multiplicity among M Dwarfs}. 1992, {\it
  \apj}, {\bf 396}, 178, DOI: 10.1086/171708

\bibitem[{{Goodwin} \& {Kroupa}(2005)}]{go_kr05}
{Goodwin}, S.~P. \& {Kroupa}, P., {Limits on the primordial stellar
  multiplicity}. 2005, {\it \aap}, {\bf 439}, 565, DOI:
  10.1051/0004-6361:20052654

\bibitem[{Heggie(1975)}]{heggie}
Heggie, D.~C., Binary evolution in stellar dynamics. 1975, {\it \mnras}, {\bf
  173}, 729, DOI: 10.1093/mnras/173.3.729

\bibitem[{{Kroupa}(1995)}]{kroupa95a}
{Kroupa}, P., {Inverse dynamical population synthesis and star formation}.
  1995, {\it \mnras}, {\bf 277}, 1491, DOI: 10.1093/mnras/277.4.1491

\bibitem[{{Kroupa}(2001)}]{kroupa}
{Kroupa}, P., {On the variation of the initial mass function}. 2001, {\it
  \mnras}, {\bf 322}, 231, DOI: 10.1046/j.1365-8711.2001.04022.x

\bibitem[{{Kroupa}(2008)}]{kroupa08}
{Kroupa}, P., {Initial Conditions for Star Clusters}. 2008, in Lecture Notes in
  Physics, Berlin Springer Verlag, Vol. {\bf  760}, {\it The Cambridge N-Body
  Lectures}, ed. S.~J. {Aarseth}, C.~A. {Tout}, \& R.~A. {Mardling}, 181

\bibitem[{K{\" u}pper {et~al.}(2011)K{\" u}pper, Maschberger, Kroupa, \&
  Baumgardt}]{mcluster}
K{\" u}pper, A. H.~W., Maschberger, T., Kroupa, P., \& Baumgardt, H., Mass
  segregation and fractal substructure in young massive clusters – I. The
  McLuster code and method calibration. 2011, {\it Monthly Notices of the Royal
  Astronomical Society}, {\bf 417}, 2300, DOI: 10.1111/j.1365-2966.2011.19412.x

\bibitem[{{Mayor} {et~al.}(1992){Mayor}, {Duquennoy}, {Halbwachs}, \&
  {Mermilliod}}]{may_etal92}
{Mayor}, M., {Duquennoy}, A., {Halbwachs}, J.~L., \& {Mermilliod}, J.~C.,
  {CORAVEL Surveys to Study Binaries of Different Masses and Ages}. 1992, in
  Astronomical Society of the Pacific Conference Series, Vol. {\bf ~32}, {\it
  IAU Colloq. 135: Complementary Approaches to Double and Multiple Star
  Research}, ed. H.~A. {McAlister} \& W.~I. {Hartkopf}, 73

\bibitem[{{Oh} {et~al.}(2015){Oh}, {Kroupa}, \& {Pflamm-Altenburg}}]{oh_etal15}
{Oh}, S., {Kroupa}, P., \& {Pflamm-Altenburg}, J., {Dependency of Dynamical
  Ejections of O Stars on the Masses of Very Young Star Clusters}. 2015, {\it
  \apj}, {\bf 805}, 92, DOI: 10.1088/0004-637X/805/2/92

\bibitem[{{Pavl{\'\i}k} {et~al.}(2019{\natexlab{a}}){Pavl{\'\i}k}, {Kroupa}, \&
  {{\v{S}}ubr}}]{pav_segr}
{Pavl{\'\i}k}, V., {Kroupa}, P., \& {{\v{S}}ubr}, L., {Do star clusters form in
  a completely mass-segregated way?} 2019{\natexlab{a}}, {\it \aap}, {\bf 626},
  A79, DOI: 10.1051/0004-6361/201834265

\bibitem[{{Pavl{\'\i}k} {et~al.}(2019{\natexlab{b}}){Pavl{\'\i}k}, {Kroupa}, \&
  {{\v{S}}ubr}}]{onc_data}
{Pavl{\'\i}k}, V., {Kroupa}, P., \& {{\v{S}}ubr}, L., {VizieR Online Data
  Catalog: ONC stars masses from literature (Pavlik+, 2019)}.
  2019{\natexlab{b}}, {\it VizieR Online Data Catalog}, J/A+A/626/A79

\bibitem[{{Plunkett} {et~al.}(2018){Plunkett}, {Fern{\'a}ndez-L{\'o}pez},
  {Arce}, {Busquet}, {Mardones}, \& {Dunham}}]{plunkett}
{Plunkett}, A.~L., {Fern{\'a}ndez-L{\'o}pez}, M., {Arce}, H.~G., {et~al.},
  {Distribution of Serpens South protostars revealed with ALMA}. 2018, {\it
  ArXiv e-prints}

\bibitem[{{Raghavan} {et~al.}(2010){Raghavan}, {McAlister}, {Henry}, {Latham},
  {Marcy}, {Mason}, {Gies}, {White}, \& {ten Brummelaar}}]{rag_etal10}
{Raghavan}, D., {McAlister}, H.~A., {Henry}, T.~J., {et~al.}, {A Survey of
  Stellar Families: Multiplicity of Solar-type Stars}. 2010, {\it The
  Astrophysical Journal Supplement Series}, {\bf 190}, 1, DOI:
  10.1088/0067-0049/190/1/1

\bibitem[{{Sana} {et~al.}(2012){Sana}, {de Mink}, {de Koter}, {Langer},
  {Evans}, {Gieles}, {Gosset}, {Izzard}, {Le Bouquin}, \&
  {Schneider}}]{sana_etal12}
{Sana}, H., {de Mink}, S.~E., {de Koter}, A., {et~al.}, {Binary Interaction
  Dominates the Evolution of Massive Stars}. 2012, {\it Science}, {\bf 337},
  444, DOI: 10.1126/science.1223344

\bibitem[{{Sana} \& {Evans}(2011)}]{sana_evans11}
{Sana}, H. \& {Evans}, C.~J., {The multiplicity of massive stars}. 2011, in IAU
  Symposium, Vol. {\bf  272}, {\it Active OB Stars: Structure, Evolution, Mass
  Loss, and Critical Limits}, ed. C.~{Neiner}, G.~{Wade}, G.~{Meynet}, \&
  G.~{Peters}, 474--485

\bibitem[{{Spitzer} \& {Hart}(1971)}]{spitzer_hart_relax}
{Spitzer}, Jr., L. \& {Hart}, M.~H., {Random Gravitational Encounters and the
  Evolution of Spherical Systems. I. Method}. 1971, {\it \apj}, {\bf 164}, 399,
  DOI: 10.1086/150855

\end{thebibliography}

\end{document}